# Treatment of gastric cancer cells with non-thermal atmospheric plasma generated in water

Zhitong Chen[a)], Li Lin, Xiaoqian Cheng, Eda Gjika, Michael Keidar[b)]

Department of Mechanical and Aerospace Engineering, The George Washington University, Washington, DC 20052, USA

[a)]Electronic mail: zhitongchen@gwu.edu
[b)]Electronic mail: keidar@gwu.edu

Non-thermal atmospheric plasma (NTAP) can be applied to living tissues and cells as a novel technology for cancer therapy. We report on a NTAP argon solution generated in deionized water (DI) water for treating human gastric cancer cells (NCI-N87). Our findings show that the plasma generated in DI water with 30-minute duration has the strongest effect in inducing apoptosis in pre-cultured human gastric cancer cells. This result can be attributed to presence of reactive oxygen species (ROS) and reactive nitrogen species (RNS) produced in water during treatment. Furthermore, the data show that elevated levels of RNS may play more significant role than ROS in the rate of gastric cancer cells death.

## I. INTRODUCTION

Gastric cancer is one of the most aggressive types of carcinomas and it is known as the second most common cause of death[1,2]. Worldwide, the number of newly diagnosed cases is predicted to reach 930,000 per year[3]. Although gastric cancer is usually managed by chemotherapy or surgery, its 5-year survival rate is approximately 15%[4]. Therefore, efforts to improve survival rates of patients with gastric cancer are one of the main challenges in current research.

Non-thermal atmospheric plasma (NTAP) can be applied to living cells and tissues due to selective cell death without influencing the heathy tissue[5-11]. The unique properties of NTAP have enabled recent biomedical applications including wood healing[12], sterilization[13], blood coagulation[14], tooth bleaching[15], skin regeneration[16] and cancer therapy[17-20]. NTAP is known for the generation of charged particles, electronically excited atoms, (Reactive Oxygen Species) ROS, (Reactive Nitrogen Species) RNS, etc[21,22]. ROS and RNS, combined or independently, are known to promote cell proliferation as well as cell death. Additionally, extreme amounts of reactive species may lead to the damage of DNA, proteins, lipids, senescence and induce apoptosis[23,24]. Recent studies showed that indirect NTAP therapy can significantly affect cancer cells[25-27]. However, there are almost no studies that report on using NTAP to treat gastric cancer cells, let alone NTAP generated in DI water.

This paper presents the effects of NTAP generated in DI water on the gastric cancer cells. NTAP device and its characterization are described, as well as the response of cancer cells to the plasma solution therapy. The voltage and current of NTAP generated in DI water were measured with a Tektronix TDS 2024B Oscilloscope. The spectra of NTAP generated in DI water were characterized by UV-visible-NIR Optical Emission Spectroscopy. The plasma density was monitored by Rayleigh Microwave Scattering system (RMS). The temperature of plasma solution was measured with FLIR Systems Thermal Imaging. The concentrations of ROS and RNS in DI water were determined by using a Fluorimetric Hydrogen Peroxide Assay Kit (Sigma-Aldrich, MO), and the Griess Reagent System (Promega, WI), respectively. The cell viability of the human gastric cancer cell line (NCI-N87) was monitored with the Cell Counting Kit 8 assay (Dojindo Molecular Technologies, MD).

## II. EXPERIMENTAL

### A. *UV-Visible Spectrum Analysis*

UV-visible-NIR was investigated on plasma discharged in water with wavelength between 200 and 850 nm. The spectrometer and the detection probe were purchased from Stellar Net Inc. In order to measure the radius of the plasma in DI water, a transparent glass plate was used to replace part of the container. The optical probe was placed at a distance of 3.5 cm in front of plasma jet nozzle. Integration time of the collecting data was set to 100 ms.

### B. *Quantification of ROS and RNS*

Fluorimetric Hydrogen Peroxide Assay Kit (Sigma-Aldrich) was used for measuring the amount of $H_2O_2$. A detailed protocol can be found on Sigma-Aldrich website. Briefly, we added 50 μl of standard curves samples, controls, and experimental samples to the 96-well flat-bottom black plates, and then added 50 μl of Master Mix to each well. The plates were incubated for 30 min at room temperature protected from light and measured emission of fluorescence by Synergy H1 Hybrid Multi-Mode Microplate Reader at Ex/Em: 540/590 nm. RNS level was determined by Griess Reagent System (Promega Corporation) according to the instructions provided by the manufacturer. Absorbance was measured at 540 nm by Synergy H1 Hybrid Multi-Mode Microplate Reader.

## *C. Cell lines*

The human gastric cancer cell line (NCI-N87) was bought from American Type Culture Collection (ATCC). ATCC is the premier global biological materials resource and standards organization whose mission focuses on the acquisition, authentication, production, preservation, development, and distribution of standard reference microorganisms, cell lines, and other materials. Cell lines were cultured in RPMI-1640 Medium (ATCC® 30-2001™) supplemented with 10% (v/v) fetal bovine serum (Atlantic Biologicals). Cultures were maintained at 37 °C in a humidified incubator containing 5% (v/v) $CO_2$.

## *D. Measurement of cell viability*

The cells were plated in 96-well flat-bottom microplates at a density of 3000 cells per well in 70 µL of complete culture medium. Confluence of each well was confirmed to be at ~40%. Cells were incubated for 24 hours to ensure proper cell adherence and stability. On day 2, 30 µl of RPMI, DI water, and plasma solutions were added to the corresponding cells. Cells were further incubated at 37 °C for another 24 and 48 hours. The viability of the gastric cancer cells was measured with Cell Counting Kit 8 assay (Dojindo Molecular Technologies, MD). The original culture medium was aspirated and 10 µL of CCK 8 reagent was added per well. The plates were incubated for 3 hours at 37 °C. The absorbance was measured at 450 nm by Synergy H1 Hybrid Multi-Mode Microplate Reader.

## III. RESULTS AND DISCUSSION

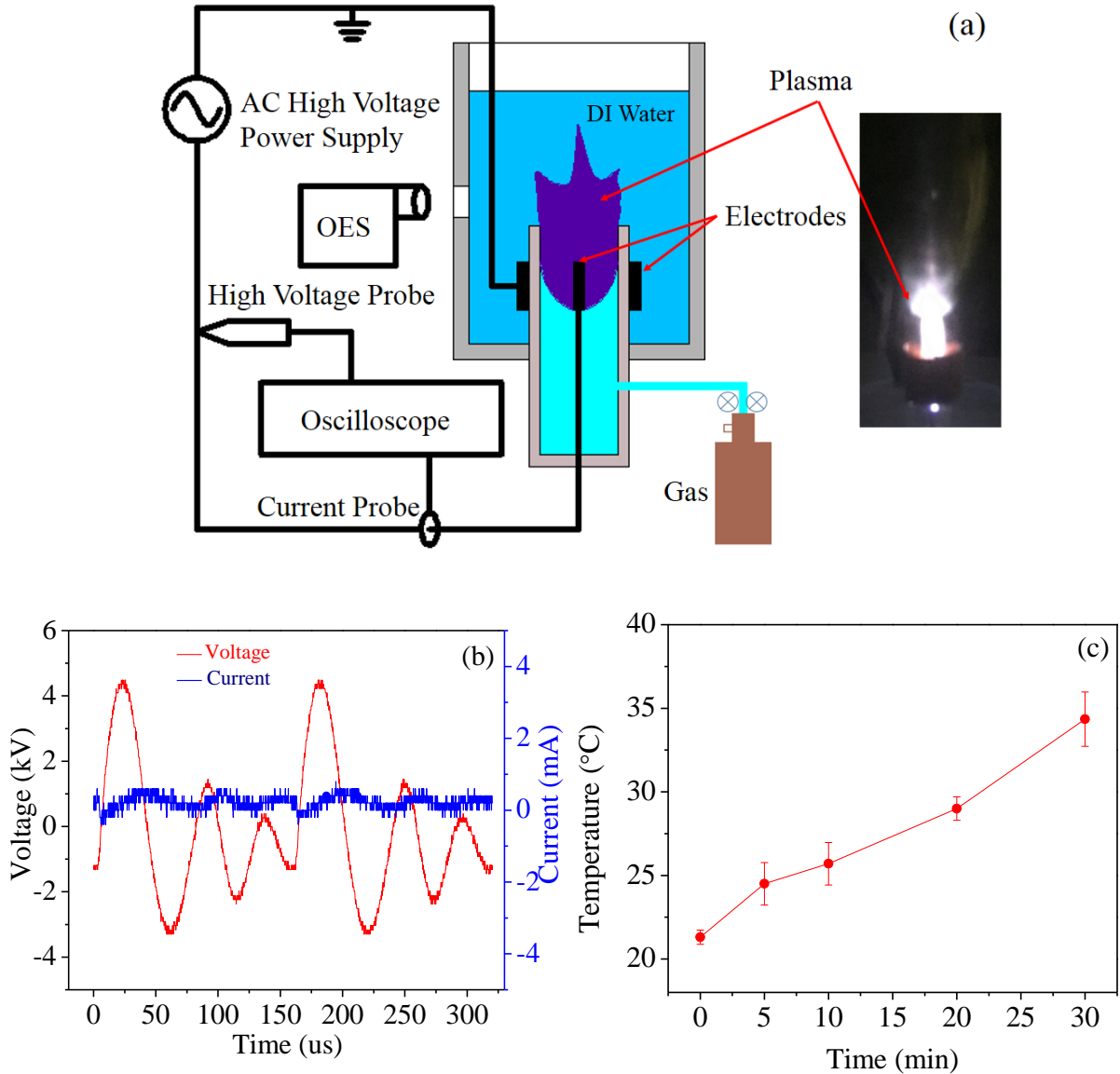

Fig. 1. Non-thermal atmospheric plasma (NTAP) generated in DI water. (a) Schematic diagram of NTAP device setup consisting of a HV pulse generator connected to a pin-to-plate electrode system submerged in DI water. (b) Voltage and current waveform. (c) Temperature changes of plasma solution with treating time increasing.

The utilized plasma device is shown in Fig.1a. Industrial grade argon with a flow rate of about 0.4 L/min was used for experiments described. The device consisted of 2 electrodes submerged in water. One electrode was a central powered electrode (1 mm in diameter) and the other one was a grounded outer electrode wrapped around the outside of a quartz tube (4.5 mm in diameter). The two electrodes were connected with a high voltage power supply. The graphs of submerged discharge current and voltage are shown in Fig. 1b, with the peak voltage about 8 kV and the average current around 0.23 mA. The frequency of the discharge generated in DI water is around 6.25 kHz. The temperature change of the plasma solutions for different treatment durations is shown in Fig. 1c, indicating that the temperature increases with treatment duration. The highest temperature increase to 34.4 ± 1.6 °C is achieved at 30 minutes' plasma treatment.

The reactive species produced by NTAP generated in DI water are shown in Fig. 2. The identification of emission lines and bands was performed according to the reference[28]. High-intensity OH/O$_3$ peak at 309 nm and low-intensity N$_2$ second-positive system ($C^3\Pi_u - B^3\Pi_g$) with its peaks at 337, 358, and 381 nm were observed. Argon lines observed in the range of 600 and 800 nm are shown in Fig. 2.

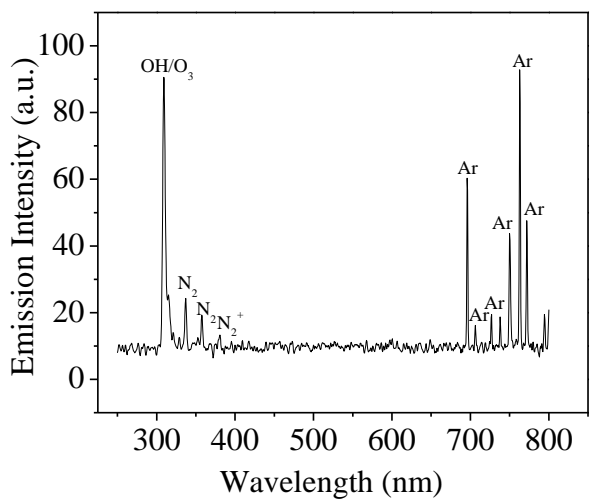

Fig. 2 Optical emission spectrum detected from the plasma submerged in DI water using UV-visible-NIR, a range of wavelength 250-850 nm

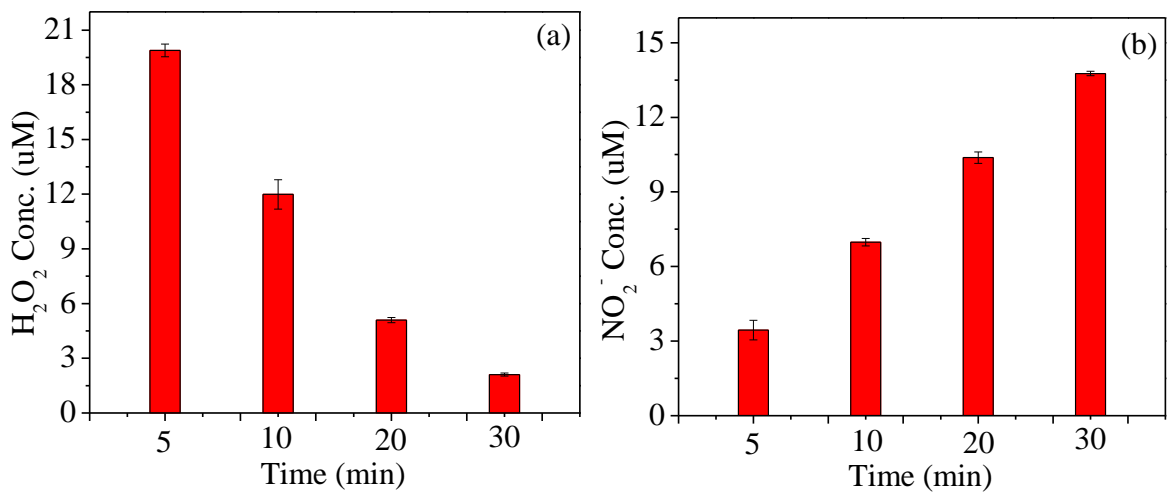

Fig. 3. $H_2O_2$ (a) and $NO_2^-$ (b) concentration in plasma stimulated DI water. $H_2O_2$ and $NO_2^-$ concentration are calculated by the concentration ratio of experimental group and control group. DI water volume is 200 ml.

NTAP can produce chemically active species in DI water. Plasma discharge produced ROS and RNS in DI water in a time-dependent manner, as shown in Fig. 3a and Fig. 3b, respectively. In order to compare the generation efficiency of ROS and RNS, we measured concentration of $H_2O_2$ and $NO_2^-$ in the case of the submerged NTAP device. $H_2O_2$ was produced in DI water within a few microseconds from hydroxyl (•OH)[29]. On the other hand, gas phase $H_2O_2$ in the afterglow also solvated into the DI water. Following mechanisms of $H_2O_2$ formation in our cases can be suggested[24,30-33].

$$Ar \rightarrow Ar^+ + e \quad (1)$$

$$Ar^+ + H_2O \rightarrow H_2O^+ + Ar \quad (2)$$

$$H_2O^+ + H_2O \rightarrow H_3O^+ + \cdot OH \quad (3)$$

$$Ar + e \rightarrow Ar^* + e \quad (5)$$

$$Ar^* + H_2O \rightarrow Ar + \cdot OH + H\cdot \quad (6)$$

$$e + H_2O \rightarrow H_2O^* + e \quad (7)$$

$$H_2O^* \rightarrow \cdot OH + H\cdot \qquad (8)$$

$$hv + H_2O \rightarrow \cdot OH + H\cdot \qquad (9)$$

$$\cdot OH + \cdot OH \rightarrow H_2O_2 \qquad (10)$$

According to Arrhenius theory[34], the decomposition rate of $H_2O_2$ increases with temperature. The temperature of the plasma solution increased with treatment time (Fig.1c), which might explain the decrease of $H_2O_2$ concentration.

Fig. 3b shows that the $NO_2^-$ concentration increases with treatment time. The $NO_2^-$ mainly originated as NO, while most of NO was formed in the gas phase during the afterglow a few milliseconds after the discharge pulse. It is known that $NO_2^-$ is a primary breakdown product of NO in DI water[35] and through the following pathways[36].

$$N_2 + e \rightarrow 2N + e \qquad (11)$$

$$N + O_2 \rightarrow NO + O \qquad (12)$$

$$4NO + O_2 + 2H_2O \rightarrow 4NO_2^- + 4H^+ \qquad (13)$$

Due to DI water contact with air, it is plausible to assume that $O_2$ and perhaps $N_2$ is coming from air. On the other hand, $N_2$ is perhaps coming from the industrial grade argon. Thus reactions (11), (12) and (13) can be used to explain the production $NO_2^-$ in argon gas

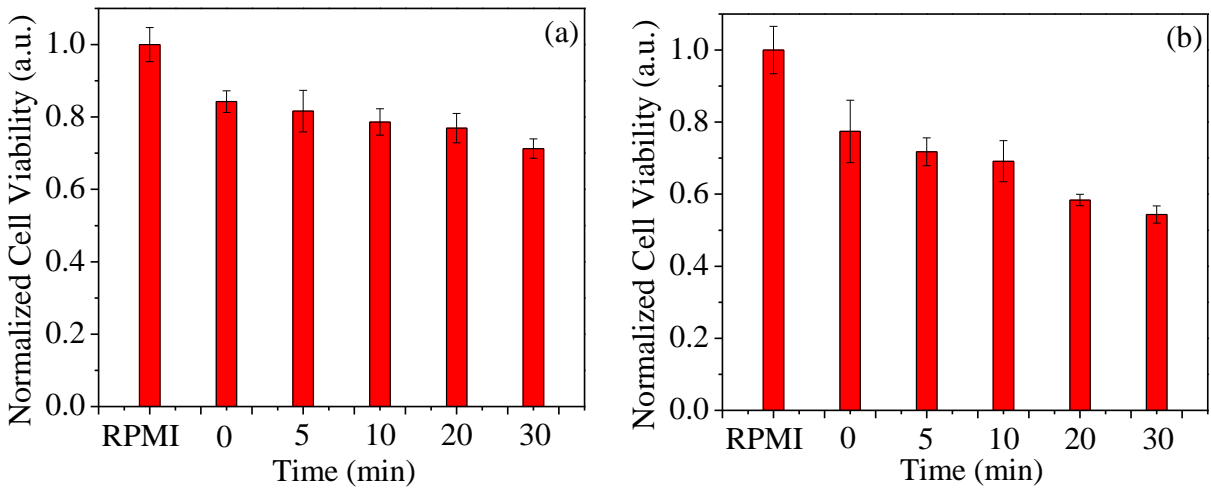

Fig. 4. The effects of the five solutions: RPMI, DI water (0 min), and plasma solutions generated in DI water during 5, 10, 20, and 30 min, on the cell viability of the human gastric cancer cells (NCI-N87). The cancer cells were plated in 96-well plates with 30 ul of plasma solutions and were incubated at 37 °C in 5% $CO_2$ for 24 hours (a) and 48 hours (b). The ratio of surviving cells from each treatment condition were calculated relative to controls.

Plasma generated in DI water was applied to gastric cancer cells. RPMI and untreated DI water were used as the control conditions. Fig. 4 shows the viability of the human gastric cancer when they were exposed to RPMI, DI water and plasma solution (5, 10, 20, and 30 mins) for 24 hours and 48 hours. At 24 hours, the viability decreased by 15.8% when the cells were treated with DI water in comparison with the RPMI control condition (Fig. 4a). The viability of cells treated by plasma solution was lower than that of the DI water and decreased with increasing treatment time. At 48 hours the viability of the cell decreased by approximately 22.6%, 28.2%, 30.9%, 41.6%, and 45.7% respectively according to treatment duration. (Fig. 4(b)). A decrease in cell viability was accompanied with an increase in the concentration of nitrite and a decrease in the concentration of $H_2O_2$. Thus, it can be seen that the strongest effect can be observed at 30 min plasma solution.

ROS and RNS are important signal mediators that regulate cell death[35]. When the cell is stimulated by environmental stress or other factors, it produces ROS that are potential signaling molecules[37]. Extreme amount of ROS in the cells may cause DNA damage, genetic instability, cellular injury and eventually induce apoptosis. RNS are pleiotropic mediators and signaling molecules involved in a large number of cell functions[38]. In some situations, RNS activate the transduction pathways causing cells apoptosis and are capable of inducing cell death via DNA double-strands break/apoptosis[39,40]. On the other hand, ROS reacts with RNS to form peroxynitrite. It leads either to caspase activation followed by apoptosis or to lipid peroxidation, protein nitration or oxidation, which can result into necrosis[41]. Our results in Fig. 3 show that the ROS concentration is highest at 5-minute treatment while the RNS concentration is highest at 30 minutes. The trend of cell death can be attributed to the increase of RNS concentration with increasing treatment time. A synergistic effect of RNS and ROS is suspected to play a key role in the apoptosis effect of plasma solution. In fact, RNS play a more important role that ROS in gastric cancer cell apoptosis under the present experimental condition.

## IV. CONCLUSIONS

In summary, non-thermal atmospheric plasma was generated in DI water using argon as a carrier gas. NTAP argon solutions were applied for treating human gastric cancer cells (NCI-N87). ROS concentration decreased with extended treatment time, while RNS concentration increased with treatment. Plasma generated in DI water during a 30-minute treatment has the strongest affect in inducing cells death. It can also be concluded that RNS plays a more significant role in gastric cancer cells death than ROS.

## ACKNOWLEDGMENTS

This work was supported in part by a National Science Foundation, grant 1465061. We thank Dr. Ka Bian and Dr. Ferid Murad from Department of Biochemistry and Molecular Medicine at The George Washington University for their help with the ROS and RNS experiments.